# Maximum Supercooling Studies in Ti$_{39.5}$Zr$_{39.5}$Ni$_{21}$ and Zr$_{80}$Pt$_{20}$ – Connecting Liquid Structure and the Nucleation Barrier

M. E. Sellers[1], D. C. Van Hoesen[1], A. K. Gangopadhyay[1], and K. F. Kelton[1]

[1]Department of Physics, Washington University in St. Louis, St. Louis, Missouri 63130, USA

Almost three quarters of a century ago, Charles Frank proposed that the deep supercooling observed in metallic liquids is due to icosahedral short-range order (ISRO), which is incompatible with the long-range order of crystal phases. Some evidence in support of this hypothesis has been published previously. However, those studies were based on a small population of maximum supercooling measurements before the onset of crystallization. Here, the results of a systematic statistical study of several hundred maximum supercooling measurements on Ti$_{39.5}$Zr$_{39.5}$Ni$_{21}$ and Zr$_{80}$Pt$_{20}$ liquids are presented. Previous X-Ray and neutron scattering studies have shown that the structures of these liquid alloys contain significant amounts of ISRO. The results presented here show a small work of critical cluster formation ($W^* = 31 - 40\ k_BT$) from the analysis of the supercooling data for the Ti$_{39.5}$Zr$_{39.5}$Ni$_{21}$ liquid, which crystallizes to a metastable icosahedral quasicrystal. A much larger value ($W^* = 60 - 99\ k_BT$) was obtained for the Zr$_{80}$Pt$_{20}$ liquid, which does not crystallize to an icosahedral quasicrystal. Taken together, these results significantly strengthen the validity of Frank's hypothesis.

## I. INTRODUCTION

Fahrenheit first noticed the tendency for liquids to exist in the liquid phase below the equilibrium melting temperature[1], also known as supercooling, which indicated the existence of a barrier to the phase transition. This barrier is now recognized as an essential ingredient to crystal nucleation—the formation of the first ordered regions in the liquid. The process of nucleation is typically understood within the framework of Classical Nucleation Theory (CNT), where the nucleation barrier arises from a competition between the thermodynamic driving free energy and the energy cost for creating an interface between the nucleating phase and the liquid. The CNT was developed by Gibbs[2], Volmer & Weber[3], and Becker & Döring[4] for gas condensation and was later extended to describe crystal nucleation in a liquid by Turnbull and Fischer[5]. Within CNT, the steady-state rate for forming crystal nuclei is expressed in terms of the atomic mobility at the crystal-liquid interface (contained in the pre-term $A^*$), the work required to form the initial critical cluster $W^*$, and the temperature $T$.

$$I^{ST} = \frac{24Dn^{*\frac{2}{3}}N_A}{\lambda^2}\left(\frac{|\Delta\mu|}{6\pi k_B Tn^*}\right)^{\frac{1}{2}}\exp\left(-\frac{W^*}{k_B T}\right) = A^*\exp\left(-\frac{W^*}{k_B T}\right) \quad (1)$$

Here, $\lambda$ is the atomic jump distance, $D$ is the diffusion coefficient, $n^*$ is the number of atoms in the critical nucleus, $N_A$ is Avogadro's number, $\Delta\mu$ is the change in chemical potential per particle, and $k_B$ is the Boltzmann constant. From CNT, $W^*$ for a spherical nucleus is:

$$W^* = \frac{16\pi}{3}\frac{\sigma^3}{\Delta g^2} \quad (2)$$

where $\Delta g$ is the Gibbs free energy difference per unit volume ($\Delta g = \Delta\mu/\bar{v}$, where $\bar{v}$ is the molecular volume) and $\sigma$ is the interfacial free energy. Turnbull and Fisher assumed that the attachment mobility at the interface was related to the diffusion coefficient in the liquid, $D$. This is generally obtained from the viscosity, $\eta$, assuming the Stokes-Einstein relation ($D = k_B T/6\pi\eta a$), where $a$ is a length that corresponds approximately to the atomic diameter.

Since CNT was originally developed for gas condensation, it does not explicitly consider the short- and medium-range order within the liquid. Frank proposed that the presence of such order, specifically icosahedral short-range order (ISRO) in metallic liquids, gives rise to the nucleation barrier in metallic liquids[6]. However, such a barrier is not limited to ISRO, any type of order that is incompatible with the long-range order in the underlying crystal phase may give rise to the nucleation barrier. Previous experimental studies have confirmed the presence of ISRO in metallic liquids[7] which was further corroborated by linking ISRO to the nucleation barrier in a $Ti_{39.5}Zr_{39.5}Ni_{21}$ supercooled liquid[8,9], providing strong support for Frank's hypothesis. However, the analysis of the supercooling data in the framework of CNT in the previous study of Ti-Zr-Ni liquids[10] was based on a small number of maximum supercooling cycles. Since nucleation is a stochastic process, as explained in section III, a quantitative analysis requires the collection of a large population of maximum supercooling temperatures. Such extensive studies are presented here for the $Ti_{39.5}Zr_{39.5}Ni_{21}$ and $Zr_{80}Pt_{20}$ liquids. The reason for choosing these alloys is that $Ti_{39.5}Zr_{39.5}Ni_{21}$ has strong ISRO in the liquid and crystallizes first to a metastable quasicrystal phase[8–10], while $Zr_{80}Pt_{20}$ contains moderate ISRO in the liquid and crystallizes to a mixture of $Zr_5Pt_3$ and $\beta Zr$ phases[11–13]. In agreement with Frank's hypothesis, $W^*$ is found to be much larger for the nucleation of the ordered phase in $Zr_{80}Pt_{20}$ than it is for the $Ti_{39.5}Zr_{39.5}Ni_{21}$ quasicrystal.

## II. EXPERIMENT

Master ingots were prepared by arc-melting high purity Zr (Smart Elements, Vienna, 99.97 at.%), Ni (Alfa Aesar 99.999 at.%), Ti (Alfa Aesar 99.999 at.%), and Pt (Alfa Aesar 99.997 at.%) on a water-cooled Cu hearth in a high purity (99.999 at.%) Ar atmosphere. A $Ti_{50}Zr_{50}$ getter located close to the sample was melted for 60 seconds prior to melting the sample to further reduce residual oxygen inside the chamber. The master ingots were melted and flipped three times to ensure chemical homogeneity, with each melting cycle lasted approximately 15 seconds to reduce mass loss due to evaporation and any residual oxygen contamination. Upon verification of negligible mass-loss (less than 0.1%), the master ingots were broken and portions were re-melted to obtain small spherical samples approximately 2.5mm in diameter, with masses ranging from 45mg to 75mg.

Nucleation can either be homogeneous or heterogeneous. Homogeneous nucleation is of most interest for these studies since it is characteristic of the thermodynamic and kinetic factors of the liquid and nucleating phases, while heterogeneous nucleation is catalyzed at specific sites, particularly by containers or sample impurities. Since metallic liquids are sensitive to oxygen and prone to heterogeneous nucleation, the studies were conducted in the high-vacuum and containerless processing environment of the Washington University Beamline Electrostatic Levitator (WU-BESL)[14,15]. The $Ti_{39.5}Zr_{39.5}Ni_{21}$ samples were processed at pressures ranging from $4.5 \times 10^{-8}$ Torr to $8 \times 10^{-8}$ Torr, while the $Zr_{80}Pt_{20}$ samples were processed at $8 \times 10^{-8}$ Torr. The spherical sample was first levitated and then melted using a high-intensity 980nm 50-Watt diode laser. A Process Sensors Metis MI18 MB8 single-color pyrometer, operating at a wavelength of 1.89μm, was used to measure the sample temperature in the 160 °C to 800 °C range. A Process Sensors Metis MQ22 two-color ratio pyrometer, operating at wavelengths of 1.4μm and 1.64μm, was used to measure sample temperatures above 600 °C. The melting plateau at 820 °C for $Ti_{39.5}Zr_{39.5}Ni_{21}$ and 1177 °C for $Zr_{80}Pt_{20}$ served as references for the calibration of the pyrometers. The temperature measurement error was corrected by assuming a constant emissivity ratio in the pyrometers; further discussion of this technique is outlined elsewhere[16]. The shear viscosity was also obtained for the liquids through the oscillating drop technique by modulating the vertical voltage to induce an oscillation in the liquid and measuring the decay time upon removing the perturbation. A detailed discussion of the density and viscosity measurement techniques can be found elsewhere[12,17].

## III. ANALYSIS AND RESULTS

Solid spherical samples were levitated, melted, and heated to several hundred degrees above the liquidus temperature to achieve maximum supercooling. The laser was then turned off and the sample cooled by radiative loss. The nucleation and growth of the crystal phase was marked by a rapid rise in the sample's temperature due to the release of the heat of fusion during crystallization, a process termed recalescence. The temperature profile during a typical heating and cooling cycle of a $Ti_{39.5}Zr_{39.5}Ni_{21}$ sample is illustrated in Fig. 1.

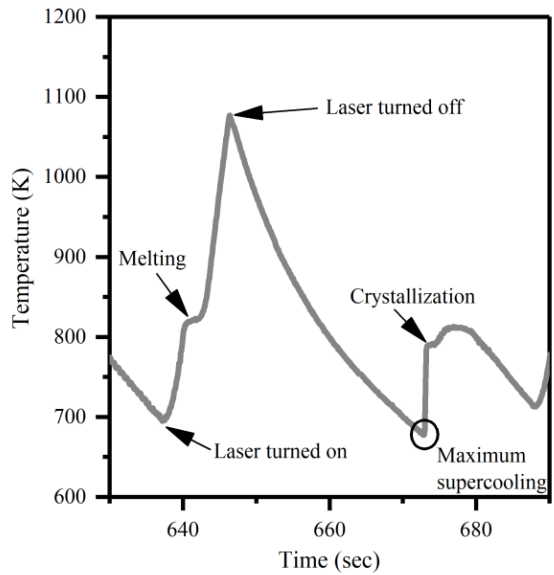

FIG 1: A representative heating and cooling cycle for $Ti_{39.5}Zr_{39.5}Ni_{21}$. In this liquid, two recalescence events are observed. The first corresponds to the transformation of the liquid to a metastable icosahedral quasicrystal phase and the second to the stable phase mixture of C14 polytetrahedral Laves phase and solid solution[9].

Such thermal cycles were repeated many times in both liquid alloys to generate a distribution of maximum supercooling temperatures ($T_u$). The first 25 cycles were excluded from the analysis because the supercooling temperatures gradually improved and stabilized around the maximum supercooling temperature with repeated processing. When consistent supercooling behavior was observed, at least 100 cycles were run and $T_u$ was measured.

To make sure that the sample quality did not change after many thermal cycles due to thermal evaporation or contamination from the residual oxygen in the BESL chamber, a consistency check was made. As seen in Fig. 2, the maximum supercooling was plotted against cycle number—a proxy for time, as each cycle happened sequentially—to check for any deterioration in supercooling. The supercooling is the difference between the liquidus temperature ($T_l$) and the maximum supercooling temperature ($T_u$), i.e. $\Delta T = T_l - T_u$. Since the icosahedral quasicrystal was the primary crystallizing phase in $Ti_{39.5}Zr_{39.5}Ni_{21}$, the supercooling for each cycle was calculated with respect to the metastable liquidus temperature of this phase (790 °C). For $Zr_{80}Pt_{20}$, the equilibrium liquidus temperature (1177 °C) was used. A linear regression analysis was performed and the $R$ value showed no significant evidence of a dependence of supercooling on cycle number ($p > 0.05$). The supercooling values are scattered evenly around the linear fit, providing evidence that the supercooling was limited by the inherent randomness of homogeneous nucleation. The introduction of contamination into the sample, such as residual oxygen in the chamber or a change in composition due to evaporation, would lead to a significant change in supercooling with time, producing a higher $R$ value, suggesting a heterogeneous nucleation mechanism.

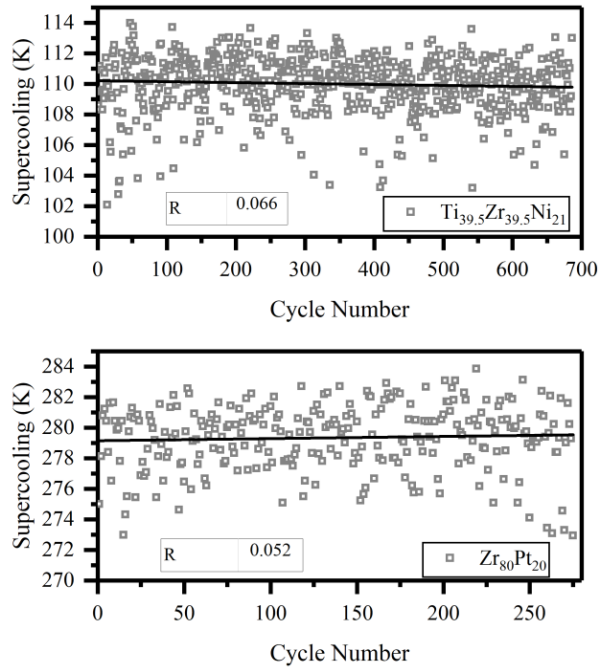

FIG 2: Maximum supercooling as a function of cycle number. No significant dependence between supercooling and cycle number is observed, indicating that the nucleation was reproducible.

The pre-factor and work of cluster formation for nucleation were obtained from these data using a statistical treatment developed by Skripov[18]. Since each nucleation event occurs independently of the others in a fixed amount of time and with an average rate, it must follow Poisson statistics. The general form of the probability density of nucleation events is:

$$\omega(T) = \frac{I^{ST}V}{Q} \exp\left(-\frac{V}{Q}\int_{T_u}^{T_l} I^{ST} dT\right) \tag{3}$$

where $I^{ST}$ is the steady-state nucleation rate $I^{ST} = A^* \exp(-W^*/k_B T)$, $A^*$ is the classical nucleation pre-term divided by the viscosity as indicated in eq. (1), $V$ is the volume of the droplet, and $Q$ is the cooling rate, $dT/dt$.

To fit the distribution of supercooling temperatures to eq. (3) and extract $A^*$, $W^*$, and $\sigma$, the driving free energy is needed as an input parameter. It was calculated by two different methods, once using the Turnbull (eq. 4.a) and again assuming the Spaepen-Turnbull[19] (eq. 4.b) approximation:

$$\Delta G = \frac{\Delta H_f \Delta T}{T_l} \tag{4.a}$$

$$\Delta G = \frac{\Delta H_f \Delta T}{T_l} \frac{2T}{T_l + T} \tag{4.b}$$

where $\Delta H_f$ is the enthalpy of fusion. The driving free energy per atom, $\Delta\mu$, is obtained by normalizing $\Delta G$ by $N_A$. The enthalpy of fusion was obtained from previously reported results for $Ti_{39.5}Zr_{39.5}Ni_{21}$ ($\Delta H_f = 8.48$ kJ/mol)[9], and estimated using Richard's Rule[20] for $Zr_{80}Pt_{20}$ ($\Delta H_f = 12.0$ kJ/mol). The interfacial free-energy $\sigma$ was also calculated twice, assuming it to be temperature independent ($\sigma$) or assuming a linear temperature dependence ($\sigma = \sigma_0 T$). The viscosity was used as an input parameter. Equation 3 was then fit to the histogram of measured $\Delta T$ values with $A^*$ and $W^*$ as fit parameters, assuming the temperature dependences of the different approximation for $\Delta\mu$ and $\sigma$. Since the different temperature dependence assumptions for $\Delta\mu$ and $\sigma$ do not change the overall shape of the fitted curve, a representative plot for each alloy is reproduced in Fig. 3 and Fig. 4. Finally, $\sigma$ was then calculated from eq. (2) and the corresponding driving free energy.

A total of 686 supercooling cycles were collected from six different $Ti_{39.5}Zr_{39.5}Ni_{21}$ samples, with masses ranging from 49.7mg to 57.5mg. The slight differences in sample volume were considered by normalizing for the different masses. A consistent supercooling temperature with an average value of 110°C (10.3% supercooling) was observed (Fig. 3) for all six samples studied; the width of the distribution is approximately 11°C. It is important to note that the distribution is asymmetric, dropping faster on the deeper supercooling side. This is typically considered to be a characteristic of homogeneous nucleation[21,22]. The average supercooling temperature was used to obtain the values for $A^*$, $W^*$ and $\sigma$ listed in Table I.

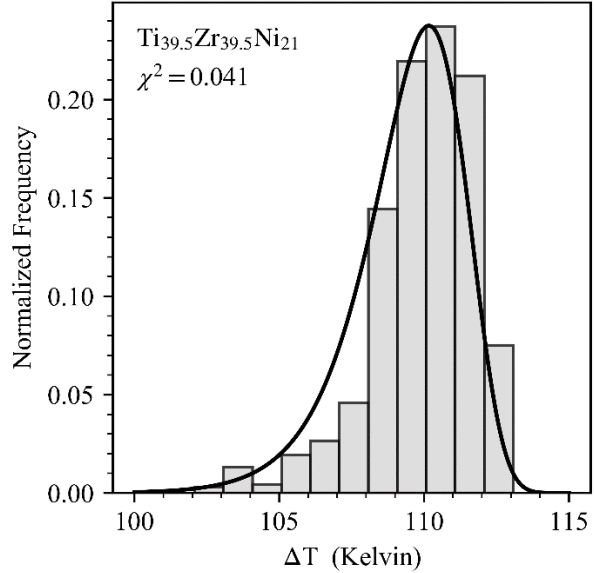

FIG 3: The histogram of maximum supercooling measurements for $Ti_{39.5}Zr_{39.5}Ni_{21}$. The average supercooling was 110 °C (10.3%). The curve is the fitted probability density, with the p-value > 0.999, indicating a good fit.

TABLE I. Calculated nucleation parameters for $Ti_{39.5}Zr_{39.5}Ni_{21}$ based on the approximation used and the assumed temperature dependence of $\sigma$.

| Approximation [a] | $A^*$ $(m^3s)^{-1}$ | $W^*/k_BT$ | $\sigma$ or $\sigma_0$ [b] |
|---|---|---|---|
| Turnbull, $\sigma \neq f(T)$ | $2.74 \times 10^{25}$ | 37.18 | 0.057 (J/m$^2$) |
| Turnbull, $\sigma \propto T$ | $7.60 \times 10^{22}$ | 31.29 | $5.68 \times 10^{-5}$ (J/m$^2$K) |
| Spaepen-Turnbull, $\sigma \neq f(T)$ | $3.72 \times 10^{26}$ | 39.79 | 0.057 (J/m$^2$) |
| Spaepen-Turnbull, $\sigma \propto T$ | $2.02 \times 10^{23}$ | 33.15 | $5.58 \times 10^{-5}$ (J/m$^2$K) |

[a] The approximation used for the driving free energy and assumption made about the dependence of $\sigma$ on temperature.
[b] The value of $\sigma$ calculated based on its assumed temperature dependence. Units given to the right of the value.

Similarly, data from 274 nucleation cycles were collected for a 72.7mg sample of $Zr_{80}Pt_{20}$, with an average supercooling of 279 °C (19.2% supercooling), as shown in Fig. 4. Compared with the measurements for $Ti_{39.5}Zr_{39.5}Ni_{21}$, the supercooling is almost two times larger and the distribution of supercooling temperatures is slightly narrower, with a width of approximately 10°C. Again, the overall shape of the fit curve does not change with the approximation assumed for $\Delta\mu$ and $\sigma$. The fit parameters from the distribution curve and the extracted nucleation properties using different approximations are listed in Table II.

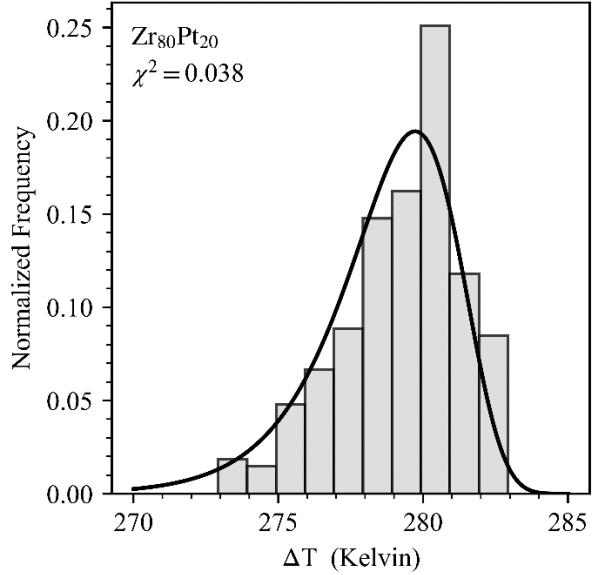

FIG 4: The histogram of maximum supercooling measurements for $Zr_{80}Pt_{20}$. The average supercooling was 279 °C (19.2%). The curve is the fitted probability density, with the p-value > 0.999, indicating a good fit.

TABLE II. Calculated nucleation parameters for $Zr_{80}Pt_{20}$ based on the approximation used and the assumed temperature dependence of $\sigma$.

| Approximation [a] | $A^*$ $(m^3s)^{-1}$ | $W^*/k_BT$ | $\sigma$ or $\sigma_0$ [b] |
|---|---|---|---|
| Turnbull, $\sigma \neq f(T)$ | $8.76 \times 10^{45}$ | 84.17 | 0.134 (J/m²) |
| Turnbull, $\sigma \propto T$ | $2.19 \times 10^{35}$ | 59.75 | $1.02 \times 10^{-4}$ (J/m²K) |
| Spaepen-Turnbull, $\sigma \neq f(T)$ | $2.44 \times 10^{52}$ | 99.01 | 0.132 (J/m²) |
| Spaepen-Turnbull, $\sigma \propto T$ | $2.91 \times 10^{38}$ | 66.95 | $9.87 \times 10^{-5}$ (J/m²K) |

[a] The approximation used for the driving free energy and assumption made about the dependence of $\sigma$ on temperature.
[b] The value of $\sigma$ calculated based on its assumed temperature dependence. Units given to the right of the value.

## IV. DISCUSSION

The general trend for $\sigma$ for the two alloy liquids is consistent with previous studies[19]. When $\sigma$ is assumed to be temperature independent, it has values ranging from 0.1 to 0.4 J/m². When $\sigma$ is assumed to be linearly dependent on temperature ($\sigma = \sigma_0 T$), $\sigma_0$ has a value near $10^{-5}$ J/m²K. From the results of the present experiment, $W^*$ and $\sigma$ appear to be insensitive to the approximation (Turnbull or Spaepen-Turnbull) used to estimate the driving free energy. This is also true when $\sigma$ is taken as temperature-independent or linearly temperature dependent (Tables I and II). However, some difference in $A^*$ is noticed under different approximations. The most important finding is that the maximum supercooling, $W^*$, and $\sigma$ are more than a factor of two larger for $Zr_{80}Pt_{20}$ than $Ti_{39.5}Zr_{39.5}Ni_{21}$ for all of the different approximations for $\Delta G$.

This is expected according to Frank's hypothesis since the supercooled $Ti_{39.5}Zr_{39.5}Ni_{21}$ liquid has significant local icosahedral order and the first nucleating solid is also a quasicrystal, as reported earlier[8,10]. The first recalescence event in Fig. 1 is due to a metastable icosahedral quasicrystal phase that subsequently transforms into a phase mixture of the polytetrahedral C14 Laves phase and solid solution[9] during the second rise after the brief plateau (Fig.1).

The $Zr_{80}Pt_{20}$ liquid and glass also contain a significant amount of local icosahedral and distorted-icosahedral order[12,23,24]. However, the crystallizing phase mixture of $Zr_5Pt_3$ and $\beta Zr$[11–13] are stable intermetallic compounds without dominant ISRO. Although the $Zr_5Pt_3$ phase contains some structural units resembling those of a quasicrystal, the simple bcc $\beta Zr$ phase does not. Since they are nucleated simultaneously, the SRO of the liquid needs to be changed during nucleation. Therefore, a larger nucleation barrier and interfacial free energy is expected for $Zr_{80}Pt_{20}$. Although stable crystal phases nucleate from the liquid for $Zr_{80}Pt_{20}$ in the present experiments with slow cooling rates, much faster melt-quenching produces metastable nano-quasicrystals[25,26]. The requirement of a faster quench rate for $Zr_{80}Pt_{20}$ to nucleate quasicrystals compared to $Ti_{39.5}Zr_{39.5}Ni_{21}$ indicates that the ISRO in the former is not so well-developed as in the latter. As a result, nucleation of the stable phases is preferred unless bypassed by a faster quench, which is consistent with the larger interfacial energy observed.

Using a different technique, Lee *et. al.* studied the nucleation of $Ti_{39.5}Zr_{39.5}Ni_{21}$[9], and observed similar values for the maximum undercooling. Since then, the method for calculating sample density has improved[17] and the statistical method outlined in this paper requires fewer assumptions for $I^{ST}$ and $A^*$ to calculate $\sigma$. As a check, the method proposed by Lee *et. al.* was also used to calculate $\sigma$. The density was first corrected, the value for $A^*$ obtained from fitting the undercooling distribution was used, the Turnbull approximation was assumed for the free energy, and a temperature-independent σ was assumed. The values obtained for $W^*/k_BT$ and $\sigma$ were 34.5 and 0.055 (J/m$^2$) respectively, which are in closer agreement with those results found from the statistical method used here. These values are smaller than those found in other alloy liquids, such as Zr-Ni[27]. This is reasonable because of the structural similarity between the liquid and primary nucleation phase in $Ti_{39.5}Zr_{39.5}Ni_{21}$; the structures are more different in Zr-Ni and $Zr_{80}Pt_{20}$.

Finally, the different magnitudes of $A^*$ need to be addressed. Based on previous studies of elemental metallic liquids performed by Turnbull, a value approximately $10^{39}$ $(m^3s)^{-1}$ for $A^*$ was proposed as a benchmark for homogeneous nucleation; smaller values, approximately $10^{29}$ $(m^3s)^{-1}$, were considered indicative of heterogeneous nucleation[28]. More recent levitation studies of pure zirconium are also consistent with this criterion[21,22,29]. For the $Zr_{80}Pt_{20}$ liquid, $A^*$ appears to be consistent with homogeneous nucleation, although it is somewhat larger when $\sigma$ is taken to be temperature independent. In contrast, the value of $A^*$ for $Ti_{39.5}Zr_{39.5}Ni_{21}$ might appear to indicate heterogeneous nucleation, since it is 13 to 17 orders of magnitude smaller. This would be surprising for several reasons. As shown in figure 2, like the case for $Zr_{80}Pt_{20}$, there was no hint of degradation of supercooling with time/cycling for $Ti_{39.5}Zr_{39.5}Ni_{21}$ during processing. The distribution of maximum supercooling temperatures is similarly asymmetric for both liquids. Finally, the samples were processed in similar high-vacuum conditions and created from the same high-purity elements. All of these observations suggest homogeneous nucleation. As pointed out in an earlier study[8], the nucleation of the i-phase in $Ti_{39.5}Zr_{39.5}Ni_{21}$ is favored in regions of the liquid that developed significant ISRO. This would give the appearance of heterogeneous nucleation and result in the lower measured pre-factor. More extensive studies in other alloy liquids are planned in the future to investigate this further.

## V. CONCLUSIONS

A systematic statistical study of hundreds of maximum supercooling measurements for $Ti_{39.5}Zr_{39.5}Ni_{21}$ and $Zr_{80}Pt_{20}$ was performed and nucleation parameters were calculated using CNT and the Skripov method. The resulting nucleation parameters were found to provide additional support for Frank's hypothesis, since the smaller values for $A^*$ and $W^*$ found in $Ti_{39.5}Zr_{39.5}Ni_{21}$ are likely due to ISRO in the liquid facilitating nucleation of the metastable quasicrystal. Meanwhile, the larger values for $A^*$ and $W^*$ found in $Zr_{80}Pt_{20}$ indicate that the structure of the liquid is different from that of the primary crystallizing phases, inhibiting their nucleation. In addition to their inherent interest for supporting Frank's hypothesis, these data will serve as a comparison for ongoing nucleation studies on the International Space Station, investigating the role of diffusion and convective stirring on nucleation, and to further understand nucleation in more complex metallic liquids.

**ACKNOWLEDGMENTS**

This material is based upon work partially supported by the National Science Foundation Graduate Research Fellowship under Grant No. DGE-1745038 and by the National Aeronautics and Space Administration Grant No. NNX16AB52G. The authors would like to thank Chris Pueblo and Robert Ashcraft for providing density and viscosity data.